\documentclass[epjST]{svjour}
\usepackage{amsmath,amssymb}
\usepackage{float}
\usepackage{imakeidx}
\usepackage[colorlinks=true,linkcolor=blue,citecolor=blue,urlcolor=magenta]{hyperref}
\usepackage[framemethod=TikZ]{mdframed}
\usepackage{footnote}
\usepackage{comment}
\usepackage{doi}
\usepackage{enumitem}
\usepackage{slashed}
\usepackage{xcolor}
\usepackage{xstring}
\usepackage{xparse}
\usepackage{graphicx}
\usepackage{subfig}
\usepackage{rotating}
\usepackage{orcidlink}
\usepackage{aastex_hack}
\usepackage{appendix}

\begin{document}
\title{A near-term quantum simulation of the transverse field Ising model hints at Glassy Dynamics}

\author{S. Ishmam Mohtashim \orcidlink{0000-0002-7382-6466}\inst{1,2,3,4}\fnmsep\thanks{\email{smohtas@ncsu.edu}} \and Arnav Das \orcidlink{0000-0003-3751-3056}\inst{5} \and Turbasu Chatterjee \orcidlink{0000-0003-4797-9124} \inst{6} \and Farhan T. Chowdhury \orcidlink{0000-0001-8229-2374}\inst{7}\fnmsep\thanks{\email{f.t.chowdhury@exeter.ac.uk}}}
\institute{
\inst{} Department of Chemistry, University of Dhaka, Dhaka 1000, Bangladesh \and
\inst{} Department of Chemistry, North Carolina State University, Raleigh, NC 27606, USA \and
\inst{} Department of ECE, North Carolina State University, Raleigh, NC 27606, USA \and
\inst{} Department of Chemistry, Purdue University, West Lafayette, IN 47907, USA \and
\inst{} A.K. Choudhury School of IT, University of Calcutta, Kolkata 700009, India \and
\inst{} Department of Computer Science, Virginia Tech, Blacksburg, Virginia 24061, USA \and
\inst{} Department of Physics, University of Exeter, Stocker Road, Exeter EX4 4QL, United Kingdom
}

\abstract{We demonstrate quantum circuit simulations of the transverse field Ising model with longitudinal fields, displaying salient features of glassy dynamics. The energy landscape and spin configurations of toy models are considered, using the Variational Quantum Eigensolver (VQE) to obtain the ground-state energies and corresponding eigenstates for a 6~$\times$~6 Ising lattice using 36 qubits and a 1-Dimensional Ising chain of length 25 using 25 qubits. The former showed disordered spin configurations for a specific mixture of values of the two fields. These insights mirror catalytic processes, where disorder within a catalyst can lead to inefficient reaction mechanisms. Results obtained from our proof-of-principle implementation make the case for kick-starting more concentrated efforts in harnessing existing quantum computational tools for computationally probing complex dynamical behaviour arising in quantum matter. Our aim is to leverage tools from quantum information processing to bring about a more nuanced understanding of the dynamics and structure of glassy systems, ultimately informing the development of novel materials and technology.}
\maketitle
\section{Introduction}\label{sec1}

Efficient means of computationally tackling large scale simulations of spin glasses are of vital importance for enabling discoveries of exotic phases of matter in extreme settings \cite{Zhang2017,Glasslike}. In addition to instigating technological advancements, the ability to probe simulations of glassy systems of realistic complexity beyond what is currently tractable via harnessing classical high-performance computing (HPC) and exascale techniques, could help us glean deep insights into the nebulous underpinnings of glassy transitions \cite{chatterjee2024features,zeng2023,Debenedetti2001,doi:10.1126/science.267.5204.1615.f} in quantum many-body physics contexts and beyond. In particular, simulated emergence of disordered phases under transverse and longitudinal fields have been validated in experimental studies on amorphous magnets and strongly correlated electron systems. Such observations highlight the role of disorder-induced phase transitions and underscore the potential of quantum simulations to aid in designing advanced materials with improved magnetic storage or catalytic performance through revealing the underlying origins of glassy characteristics. Remarkably, glassy behaviour also manifests in systems atypical \cite{PhysRevResearch.6.013093,Chiang_2016,Sibani_1993,PhysRevLett.71.1482,PhysRevA.28.2408} of quantum matter, including, but not limited to, in complex biological systems and in meta-community models described by statistical mechanics \cite{PhysRevB.105.094409,PhysRevE.103.062403,PhysRevResearch.1.032038}. 

In this work, we explore the prospects of harnessing digital quantum simulation (DQS) techniques for uncovering otherwise inaccessible novel phenomena in disordered spin configurations of a square lattice Ising model with transverse and longitudinal fields. A hybrid quantum algorithm devised for simulating in particular the ground state energies of spin Hamiltonians, a crucial problem in quantum chemistry and condensed matter physics, is the Variational Quantum Eigensolver (VQE) \cite{Peruzzo_2014,10313751,VQEspin}. We choose this technique since it was developed taking into account the low-circuit depths and relatively high noise rates of existing quantum hardware platforms \cite{singkanipa2025beyond,zimboras2025myths,aplQ,govindaraja2024fidelity,Moll_2018,McClean_2016,Peruzzo_2014}. Classical computational approaches, such as molecular dynamics and Monte Carlo simulations, face significant challenges in managing the high computational complexity involved in simulating large glassy systems over extended periods \cite{lee2021quantum,Sachdev_2011,Landau_Binder_2009,10.1063/5.0173591}. In particular, they struggle with accuracy when modeling the complex electronic wavefunctions of many-electron systems, as these wavefunctions grow exponentially and become computationally intensive to handle precisely \cite{prxL,prxL2,Landau_Binder_2009,RevModPhys.73.33}. The VQE addresses this challenge by enabling the modeling of these complex wavefunctions in polynomial time, significantly reducing the computational costs.

There is a lack of extensive literature on simulating glassy systems using quantum computational techniques, which we seek to address. The transverse field Ising model (TFIM) with longitudinal field \cite{Wang_1994,PFEUTY197079} exhibits disordered spin systems \cite{PhysRevE.99.012122,Simon2011}, where glassy dynamics have been found to be emergent \cite{Schultzen_2022}. The glassy behaviour in these models can be attributed to quenched spin disorder and relaxed spin-flip dynamics \cite{Safavi_Naini_2019}, and the disordered phase in these models can in fact help model many near-optimal solutions to combinatorial optimization problems, neural networks and minority games \cite{10.3389/fphy.2014.00005,10.2307/j.ctt12f4hf}. These solutions refer to the local minima in the energy landscape, thereby indicating meta-stability in its phases \cite{yan2016dynamics}. Glassy systems present rugged energy landscapes characterized by numerous local minima, complicating the identification of the global minimum and the accurate simulation of their dynamics \cite{GlassCrystal}. Their dynamics involve extremely long relaxation timescales, which require extensive computational resources to simulate over practical timescales. 

\begin{figure*}[t!]
    \centering
    \includegraphics[width=0.9\linewidth]{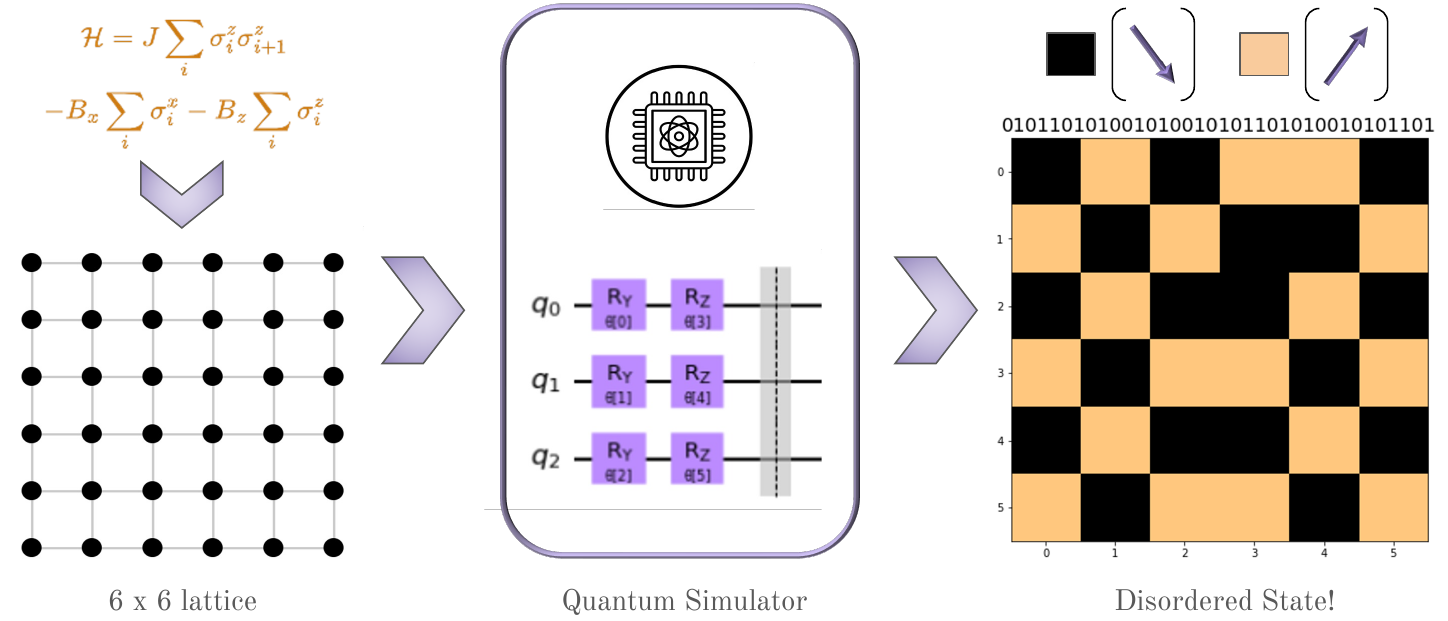}
    \caption{A schematic illustrating the proof-of-principle implementation.}
    \label{fig:6x6-3d-lattice}
\vspace{-1.5em}
\end{figure*}

The inherent disorder and heterogeneity of glasses add further complexity to modeling efforts, as they demand meticulous attention to local variations in structure and dynamics \cite{annurev:/content/journals/10.1146/annurev.physchem.58.032806.104653}. Moreover, glasses typically reside in non-equilibrium states, which require advanced simulation techniques to accurately capture their behavior over time, particularly during state transitions \cite{cugliandolo2002dynamicsglassysystems}. Scrutinising their underlying quantum spin systems \cite{gao2022emergence} represent a cornerstone in the study of condensed matter physics due to their rich emergent behaviors. This could play a pivotal role in advancing our understanding of quantum phase transitions \cite{Chen2024,Chen2019,QSL}, and potentially informing the discovery of exotic material properties in extreme conditions. Our motivation is to push the boundaries of NISQ era computing applications by simulating the phases of spin glasses \cite{Altieri_2024,PhysRevB.105.094409,optical}, which is why we propose the incorporation of a future Qiskit \cite{Qiskit} module that formulates problem Hamiltonians of interest based on the number of spins and the geometry of a lattice encoding, and furthermore conduct benchmarking of an appropriately reduced model on IBM quantum hardware.
\vspace{-1em}
\section{Methods}\label{sec2}

To simulate glassy dynamics on a quantum computer, we have primarily consider the Ising model with transverse and longitudinal fields, with the spins having nearest neighbor interactions along the horizontal and vertical directions. Hence, the interactions along the diagonal directions are not present in our model. The ground state energies of the models were found using VQE \cite{Peruzzo_2014}.

\vspace{-1em}
\subsection*{Variational Quantum Eigensolver}
\vspace{-0.5em}
VQE builds on the variational principle of quantum mechanics, which states that for any Hamiltonian $H$ and any trial wave function $|{\psi(\boldsymbol{\theta})}\rangle$ parameterized by variables $\boldsymbol{\theta}$, the expectation value of $H$ is denoted by $\langle{\psi(\boldsymbol{\theta})|H|\psi(\boldsymbol{\theta})}\rangle$. This is expressed by the inequality
$E_0 \leq \langle{\psi(\boldsymbol{\theta})|H|\psi(\boldsymbol{\theta})\rangle}$, where $E_0$ is the lowest eigenvalue of $H$. Succinctly, one can describe the steps in VQE as follows: 
\begin{enumerate}
    \item Prepare ansätze $|{\psi(\theta)}\rangle$ on a quantum computer.
    \item Measure expectation value $H(\theta) = \langle{\psi(\theta)|H|\psi(\theta)}\rangle$.
    \item Based on the measurements, use a classical optimiser to vary \\
    the $\theta$ such that $\langle{\psi(\theta)|H|\psi(\theta)}\rangle$ becomes smaller till it converges to the minimum value.
\end{enumerate}
The algorithm iteratively optimises the parameters $\boldsymbol{\theta}$, a crucial step in fact performed classically, which is why it is described as a hybrid-quantum \cite{aparicio2023minimizing} algorithm. In essence, the circuits run natively on quantum hardware, but with circuit parameters optimised on a classical computer.

\subsection*{TFIM model with Longitudinal Field}

The Hamiltonian \( \mathcal{H} \) of the TFIM model in the presence of a longitudinal field is expressed using tensor products of Pauli operators. It encapsulates the energy of a system of spins arranged on a lattice and is given by \cite{PhysRevE.99.012122}
\begin{equation}
    \mathcal{H} = J \sum_{i} \sigma_{i}^{z} \sigma_{i+1}^{z} - B_x \sum_{i} \sigma_{i}^{x} - B_z \sum_{i} \sigma_{i}^{z}, 
\end{equation}
where the Pauli operators \( \sigma^x \) and \( \sigma^z \) represent the $x$ and $z$ components of the spin, respectively. The coupling constant \( J \) governs the strength of the spin-spin interaction along the $z$-axis, with a ferromagnetic tendency indicated by a negative \( J \). The term \( -B_x \sum_{i} \sigma^x_i \) corresponds to the transverse magnetic field which induces quantum fluctuations, leading to spin flips in the $x$ direction. On the other hand, \( -B_z \sum_{i} \sigma^z_i \) represents the longitudinal magnetic field's effect on the alignment of spins along the $z$-axis. In essence, each spin is mapped onto a qubit, allowing for the representation of spins in superposition states \cite{Lucas_2014}. The the value of \( J \) is standardized to 1, thus normalizing the energy scales associated with the spin interactions and magnetic field effects \cite{optical}.

For the 1D case, we considered an Ising chain of length $n=25$. The Hamiltonian $\mathcal{H}$ is constructed using a custom-coded utility that prints it out using elementary Pauli operators \cite{PFEUTY197079}. This 25 qubit Hamiltonian is then fed into the VQE protocol. The specific trial wavefunction, or ansätze, employed in this implementation of the VQE algorithm is known as the \textit{TwoLocal} ansätze. This ansätze comprises two types of rotation blocks, \( R_x \) and \( R_y \), which correspond to rotations about the $x$ and $y$ axes on the Bloch sphere, respectively. In this setup, the ansätze is intentionally designed without entanglement between qubits \cite{ansatz}. While entanglement is a quintessential quantum resource that frequently enables quantum algorithms to surpass classical performance, certain problem structures may be adequately addressed with a non-entangled \cite{chowdhury24,mouloudakis2023interspecies,PhysRevA.98.032309,PhysRevA.109.022618} ansätze without strictly quantum correlations, thus reducing the resources required to simulate the quantum circuit, but not necessarily to the extent that it becomes readily simulable classically. Optimization of ansätze parameters is performed using the \texttt{COBYLA} optimiser, a numerical optimization method that is particularly apt for scenarios with noisy or computationally intensive objective functions, common in quantum circuit evaluations \cite{Pellow_Jarman_2021}. The optimiser is set to perform up to 25,000 iterations, meticulously adjusting the ansätze parameters to minimize the expectation value of the Hamiltonian, to converge to the system's ground state energy \cite{gibbs2025exploiting,QCircuitLearning}. Computational resources required to implement VQE are considerable, as the state space scales exponentially with the number of qubits. To manage this, we use the \texttt{Qiskit Aer\_gpu} backend, optimised for execution on Graphics Processing Units (GPUs) \cite{Qiskit,Jones2019,McClean_2016}.

\begin{figure}
    \centering
    \includegraphics[width=0.47\linewidth]{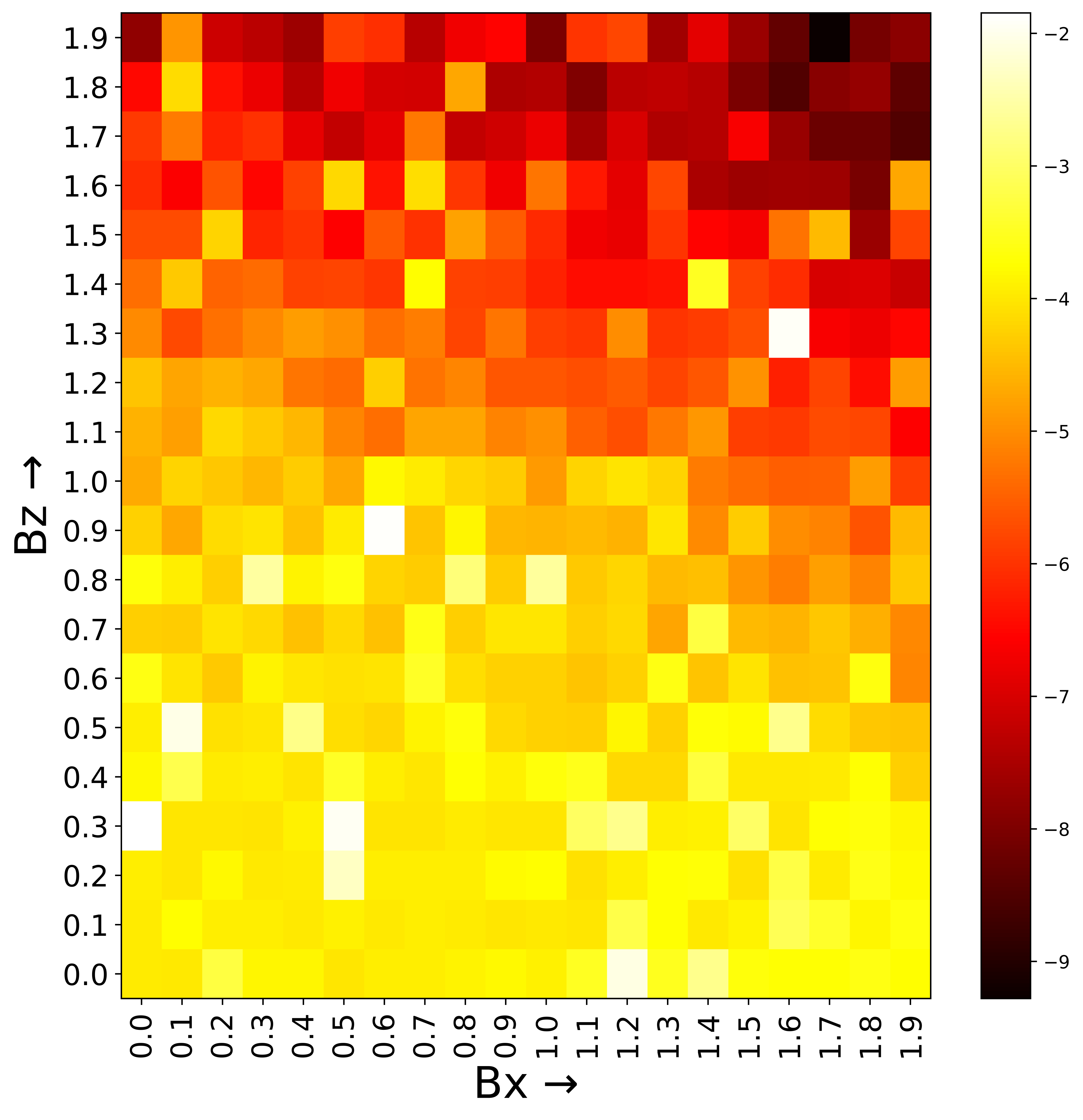}
    \vspace{-1em}
    \caption{Energy Landscape for 1D Ising for varying $B_x$ and $B_z$.}
    \label{heatmap}
    \vspace{-1em}
\end{figure}

\begin{figure*}[t]
\centering
\subfloat[$B_z=1, B_x=0.5$]{\includegraphics[width = 0.25\textwidth]{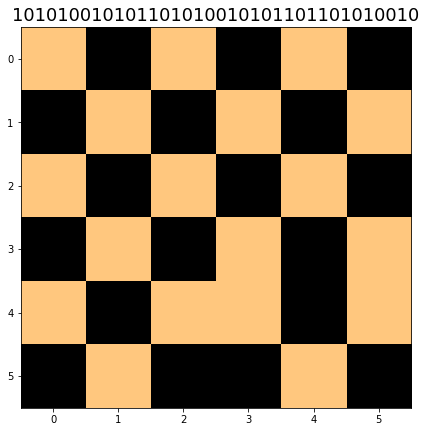}}
\hfill
\subfloat[$B_z=1.5, B_x=0.25$]{\includegraphics[width = 0.25\textwidth]{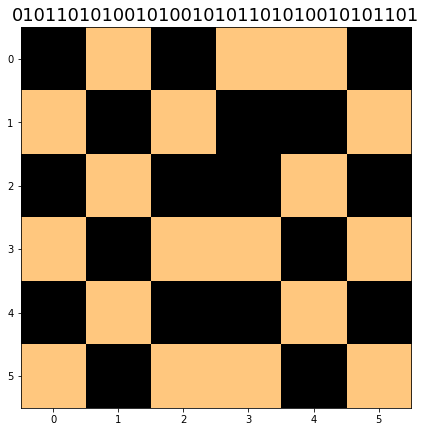}}
\hfill
\subfloat[$B_z=0, B_x=2.0$]{\includegraphics[width = 0.25\textwidth]{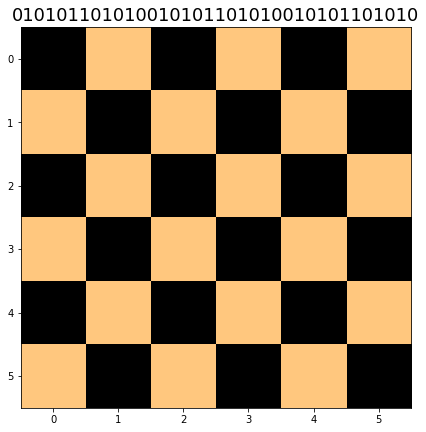}}
\\[1em] 
\subfloat[$B_z=1, B_x=2.0$]{\includegraphics[width = 0.25\textwidth]{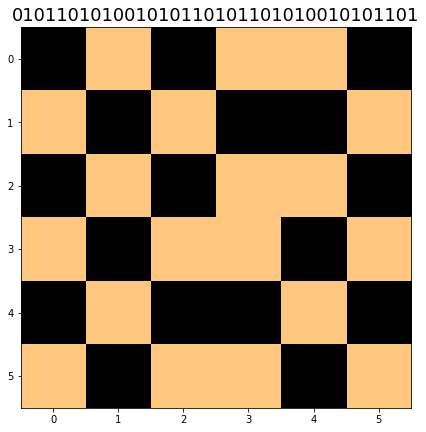}}
\hspace{2em} 
\subfloat[$B_z=1.5, B_x=1.5$]{\includegraphics[width = 0.25\textwidth]{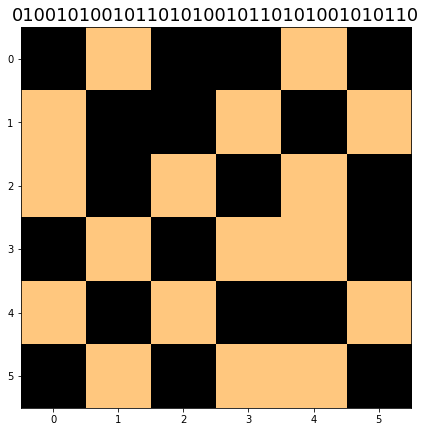}}
\caption{2D Ising plots with $J=1$ and varying $B_x$ and $B_z$.}
\label{2d}
\vspace{-1em}
\end{figure*}

We look at the 2-Dimensional Ising model as the behavior of the entire system changes due to the variations in the number of neighbors and the interactions between spins. \cite{PhysRevB.110.155128,PhysRev.65.117} For spins on the corner of the 2D lattice, we have two interactions. Spins on each edge have three interacting spins. In general, spins throughout the lattice have 4 interacting spins, when considering the 2D case. The Ising model was constructed on a square lattice of $6\times6$.
The Hamiltonian $\mathcal{H}$ is so constructed using a custom-coded utility that prints out the Hamiltonian using elementary Pauli operators. This 36-qubit Hamiltonian is then fed into a Variational Quantum Eigensolver algorithm, that uses a TwoLocal ansätze. The ansätze uses $R_x$ and $R_y$ rotation blocks with no entanglement in them. The VQE used a \texttt{COBYLA} optimiser whose maximum iterations were set at 25,000 iterations. Though most simulation runs capped out at a few thousand iterations. We used the \texttt{matrix\_product\_state} \cite{chen2020effects} method on the \texttt{AerSimulator} backend to simulate the VQE algorithm. To construct the 2D transverse field Ising model Hamiltonian with longitudinal fields, we developed a custom package, \textbf{Qiskit Glassy Dynamics}, to streamline Hamiltonian generation for quantum simulations. The Hamiltonian, given by 
\[
H_{2D} = J \sum_{\langle i,j \rangle} \sigma^z_i \sigma^z_j - B_x \sum_{i} \sigma^x_i - B_z \sum_{i} \sigma^z_i,
\]
was implemented as a combination of interaction terms (\( \sigma^z \sigma^z \)) and field terms (\( \sigma^z \), \( \sigma^x \)). Using this package, we defined the Hamiltonian components with efficient syntax: the nearest-neighbor interaction term was created with \texttt{Ising2DHamiltonian((6, 6))}, while the longitudinal and transverse field terms were generated using \texttt{FieldHamiltonian((6, 6), 'Z')} and \texttt{FieldHamiltonian((6, 6), 'X')}, respectively. The final Hamiltonian was assembled as a weighted sum of these terms.
\vspace{-1em}
\subsection*{Choice of Ansätze in Variational Quantum Algorithms}

In variational quantum algorithms, selecting an appropriate ansätze is crucial for balancing expressivity, optimization efficiency, and hardware feasibility. Our ansätze, composed of \( R_x \) and \( R_y \) rotations, is particularly well-suited for models with competing interaction terms, such as the transverse-field Ising model (TFIM) with longitudinal field. Our Hamiltonian consists of \( \sigma^z \sigma^z \) interaction terms and a transverse field in the \( x \)-direction and a longitudinal field in the \( z \) direction,  making \( R_x \) and \( R_y \) rotations a natural choice for capturing quantum fluctuations while maintaining physical interpretability. These rotations efficiently span both the computational (\( \sigma^z \)-basis) and superposition (\( \sigma^x \)-basis) states, ensuring that the ansätze aligns with the structure of the Hamiltonian. This alignment allows for efficient representation of both ground and excited states while avoiding excessive parameterization, which can lead to optimization difficulties. Additionally, problem-inspired ansätze such as ours enhance trainability by reducing the likelihood of barren plateaus, which commonly arise in deep circuits.  

It is pertinent to discuss the potential of hardware-efficient ansätze as an alternative, since they promise broader expressivity \cite{Leone_2024}. However, such ansätze often also introduce significant circuit depth due to additional entangling layers, leading to increased susceptibility to noise and decoherence effects on near-term quantum devices . Since near-term hardware is limited by gate fidelity and coherence times, ansätze with excessive entangling operations may not always be practical. Our \( R_x \)-\( R_y \) ansätze mitigates this issue by reducing unnecessary entangling layers while maintaining sufficient expressivity, providing an optimal balance between accuracy and noise resilience. Compared to hardware-efficient ansätze, our approach is more physically relevant to the problem, results in a more stable optimization landscape, and is less prone to trainability issues. 

\section{Discussion}\label{sec3}

We observe in the energy landscape generated in Fig. \ref{heatmap}, that an increase in the longitudinal field leads to a lower energy of the Hamiltonian, but an increase in the transverse field does not lower the energy as much. The role of the transverse field is primarily to induce quantum effects \cite{PFEUTY197079,PhysRevResearch.1.033141} which give rise to alternating ferromagnetic and anti-ferromagnetic interactions, which in turn produce the disordered phase. Thus, the transverse field is essential for disordered spin configuration to arise \cite{PhysRevB.53.8486}. The energy landscape, which graphically represents the energy of a system as a function of its states or configurations \cite{CompMag2}, is essential for understanding the evolution of the system and the stable states into which it may settle \cite{Bapst_2013}. The longitudinal field is aligned with the direction of the spins. An increase in this field tends to align the spins in its direction, thus reducing the system's energy. This is because the spins naturally align with an external magnetic field, a state that typically corresponds to a lower energy configuration \cite{CompMag}. This effect can be visualized as a deepening of the energy well in the landscape, indicating where the system's ground state is likely to be \cite{2013LNP...862.....S}. Conversely, the transverse field is oriented perpendicular to the direction of the spins. An increase in the transverse field does not reduce the energy as significantly because it competes with the spin alignment, introducing quantum fluctuations due to its perpendicular orientation \cite{QA}. These fluctuations can cause the spins to flip, thereby introducing disorder into the system \cite{RBStinchcombe_1973,1968JPSJ...24...51S}.

The most notable thing to see in the 2D case is that the three most probable configurations for each case are nearly identical: a pattern arises in simulations, indicating that a coherent study is possible. We can now see the disordered phase region, which most published literature does not account for \cite{PhysRevE.90.032101}. The spin configurations with the highest counts/probabilities of occurrence are shown in Fig. \ref{2d}, with the rest reported in a repository \cite{das2024quglassyising}. The observation that the three most probable spin configurations are almost identical suggests a symmetrical or patterned order in the system. This implies that even complex quantum systems can exhibit some degree of predictable behavior \cite{Balents2010}. Recognizing these patterns is crucial for deciphering the behaviors of quantum many-body systems and can be pivotal in engineering materials with tailored magnetic characteristics \cite{Hastings_2007}. When $B_z=0$ and $B_x=2$, we can see a perfect anti-ferromagnetic spin configuration in the absence of the longitudinal field. The configuration with the $2^{nd}$ and $3^{rd}$ highest counts shows near-perfect anti-ferromagnetism as well. We hypothesize that a longitudinal field is essential for a disorder to occur. We define the disordered phase as the spin configurations when a consistent stretch of spin configuration shows an anti-ferromagnetic phase and a consistent stretch of spins show a ferromagnetic phase \cite{PhysRevE.99.012122}. Thus the disordered phase is basically a combination of consistent and semi-symmetric ferromagnetic and anti-ferromagnetic phases in spin configuration. This is seen in the spin configurations when $B_z=1$ and $B_x=2$, where we see a mixture of anti-ferromagnetic and ferromagnetic spin configurations, i.e., a disordered state. The ferromagnetism is seen in local moments. If we compare cases $B_z=1$, $B_x=0.5$ with $B_z=1$, $B_x=2$, we see that for the same values of $B_z$, an increase in disorder occurs with the increasing $B_x$. $B_z=1, B_x=0.5$, shows mostly anti-ferromagnetism, whereas $B_z=1, B_x=2$, develops local moments of ferromagnetism. $B_z=1.5, B_x=1.5$ shows a  relatively more paramagnetic state i.e., spins are more random. 

Further increase of $B_z$, in theory, should increase the randomness. Decreasing $B_x$, $B_z=1.5$, $B_x=0.25$, shows a disordered state \cite{PhysRevB.105.L020201}. Therefore, we observe a disordered state only for a specific combination of values of the transverse and longitudinal Ising fields, i.e., the disordered phases occupy a narrow region between the anti-ferromagnetic phase and the paramagnetic phase \cite{PhysRevE.63.016112}. The role of the longitudinal magnetic field in a quantum spin system is to align the spins along its direction \cite{doi:10.1126/sciadv.abc5511}. When only a transverse magnetic field is present, spins are influenced to orient perpendicularly to the field, leading to a certain degree of order that minimizes the system's energy in that configuration. However, the introduction of a longitudinal field adds a competing interaction \cite{pattern}. Instead of the spins aligning solely in response to the transverse field, they now also tend to align with the longitudinal field. This competition between the fields disrupts the previously established order. Spins that were once potentially in a uniform state due to the transverse field now experience a tug-of-war as the longitudinal field encourages alignment in a different direction \cite{10.1143/PTP.46.1337}. The result is a more complex arrangement of spins, with some following the transverse field and others aligned with the longitudinal field, leading to a mixed or disordered phase \cite{RBStinchcombe_1973}. The interplay between these fields can result in a rich tapestry of spin configurations, with regions exhibiting different kinds of magnetic ordering or even a lack of order entirely \cite{Nishimori_1980}. Such disruption is not merely a perturbation but is fundamental to the system's behavior, as it can lead to phase transitions where the material's properties change abruptly and significantly \cite{PhysRevLett.77.940}. In the study of quantum criticality \cite{10.1063/1.3554314}, understanding how the longitudinal field disrupts the order is vital. It is at the heart of many phenomena in condensed matter physics, such as the emergence of exotic states of matter, and has practical implications for the development of new quantum technologies, including quantum computation, metrology, and magnetic storage devices \cite{doi:10.1021/acs.jpclett.2c02840,9928639,Chen2019,1994PhR...239..179C}.

\begin{figure}
    \centering
    \includegraphics[width=0.8\linewidth]{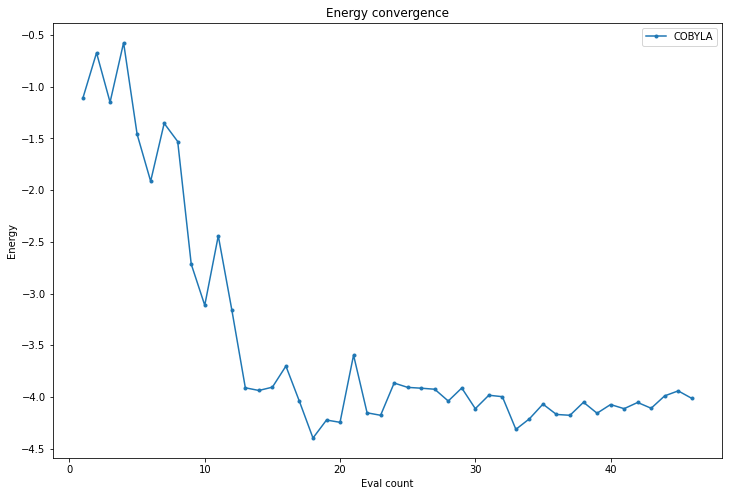}
    \caption{Energy Convergence for 1D 7-spin Ising model.}
    \label{real_deviceOslo}
    \vspace{-1em}
\end{figure}

\vspace{-1em}
\section{Custom Qiskit package and run on IBM-Q Device}\label{sec4}

We propose the incorporation of a dynamic, modular package \texttt{QuGlassyIsing} for glassy dynamics that interfaces with Qiskit \cite{Qiskit}. The central objective of \texttt{QuGlassyIsing} is to provide a user-friendly interface for constructing Hamiltonians associated with the TFIM model, incorporating longitudinal fields. By giving precedence to constructs that are accessible to users, our objective is to provide researchers with the tools to efficiently traverse the complex landscape of glassy dynamics, and facilitate finding more intuitive connections for researchers studying the quantum behavior underlying disordered phases \cite{mildner2023topological,Schiulaz_2014,PhysRevB.51.6411,refId0}. One of the core functions of the package is \texttt{Ising2DHamiltonian}, which constructs the nearest-neighbor interaction Hamiltonian for a 2D lattice. This function iterates over all lattice points in the specified grid, identifies their horizontal and vertical neighbors, and constructs interaction terms for each pair of neighbors. These terms are represented as Pauli strings (\( \sigma^z \sigma^z \)) and summed to form the overall interaction Hamiltonian. The function also handles edge cases, ensuring that boundary points with fewer neighbors are correctly accounted for. The output is a Hamiltonian that encodes all nearest-neighbor interactions within the lattice, ready for direct use in quantum simulations.

The package also includes the \texttt{FieldHamiltonian} function, which generates field terms for longitudinal (\( \sigma^z \)) or transverse (\( \sigma^x \)) fields. This function constructs the Hamiltonian by applying the specified Pauli operator to each spin in the lattice. It supports arbitrary lattice dimensions, making it adaptable to both 1D and 2D systems. By separating the interaction and field terms into modular components, the package allows researchers to easily customize the Hamiltonian for different scenarios, such as varying field strengths or interaction geometries. Another key feature of the package is the \texttt{get\_interaction} function, which defines the interaction between two specific qubits in the lattice. This function verifies the validity of the lattice points, constructs the interaction term by applying \( \sigma^z \) to the two qubits and identity operators (\( I \)) to all others, and returns the result as a properly formatted Pauli string. It serves as the building block for \texttt{Ising2DHamiltonian}, ensuring that all nearest-neighbor interactions are accurately represented in the final Hamiltonian. Together, these functions streamline the process of constructing Ising model Hamiltonians. The modular design of the package not only simplifies the generation of Hamiltonians but also ensures that the methodology can be extended to larger systems or alternative lattice geometries, such as triangular or hexagonal lattices. By leveraging this package, we efficiently constructed the Hamiltonian for a \(6 \times 6\) lattice (36 qubits), combining interaction terms and field contributions to simulate the system dynamics. This approach highlights the versatility and practicality of the Qiskit Glassy Dynamics package for exploring complex quantum systems. As of writing this article, \texttt{QuGlassyIsing} has a release candidate \texttt{v0.0.1-alpha}. Our emphasis on formulating a concise syntax within \texttt{QuGlassyIsing} is driven by the motivation that an efficient interface enhances a researcher's capacity to expedite intricate simulations. 

To gauge how well the results can be appropriately replicated on a real quantum device of the NISQ era \cite{Shi2025602,trev_ieee}, we implemented a reduced 7 spin Ising model with unit longitudinal and unit transverse field. On IBM-Q's 7 qubit Oslo device, setting the interaction strength and field strength to unit length to simplify the simulation, the energy convergence profile exhibited more fluctuations and a slower convergence rate. As seen from Fig. \ref{real_deviceOslo}, the final energy value was higher than that of the theoretical minimum. The plot highlights the fluctuations in the energy convergence due to noise and hardware-induced errors which include gate errors, readout errors, and qubit decoherence. The run is to be treated strictly as a proof-of-concept. Further details are presented in the Appendix.

\section{Conclusion}\label{sec5}

We have shown that probing toy models emulated using a near-term digital quantum simulator, allows for detecting disordered phases in one-dimensional (1D) and two-dimensional (2D) Ising models. Specifically, we observed that the introduction of a longitudinal field in the Ising models plays a crucial role in developing regions of disorder, interspersed between states of anti-ferromagnetic and ferromagnetic order. We have observed the importance of the longitudinal field in the Ising models for developing regions of disorder, between states of anti-ferromagnetic and ferromagnetic states. Overall, our findings suggest the formation of glassy relaxation in these spin systems. The findings in this work align closely with experimental and theoretical studies on spin glasses \cite{Vincent_2018}, particularly in disordered materials such as Fe–Mn–Al alloys, \cite{10.1063/1.364458} amorphous magnets \cite{Kolesnik_2003}, and nanostructured systems \cite{10.1063/1.4881498}. The simulated disordered phases, characterized by transitions between ferromagnetic and antiferromagnetic configurations, parallel the competing interactions and frustration-induced spin-glass behavior observed experimentally. Furthermore, the simulated narrow regions of disorder between antiferromagnetic and paramagnetic phases align with experimental observations of nanoscale variations influencing spin behavior and phase transitions in nanostructured alloys \cite{10.1063/1.4881498}. Furthermore, spin configurations play a vital role in catalysis, where the alignment or disorder of spins directly influences catalytic performance. Ordered spin configurations, particularly when subjected to a magnetic field, enhance catalytic activity by enabling efficient electron transfer, a key process in catalysis. In contrast, disordered spin states disrupt these processes, resulting in lower catalytic efficiency. 

Near-term intermediate-scale quantum (NISQ) devices are limited by short coherence times, gate errors, and restricted qubit connectivity, constraining the depth and accuracy of quantum simulations \cite{lall2025}. While our approach successfully leverages variational algorithms to probe glassy dynamics, these limitations prevent scaling to larger and more complex systems. Fault-tolerant quantum computing (FTQC) with quantum error correction (QEC) will enable deeper circuits and more precise state evolution, allowing for accurate exploration of long-time dynamics, phase transitions, and emergent behavior in disordered quantum systems. As QEC improves hardware fidelity \cite{malcolm2025}, our methods will scale to study larger spin lattices and more intricate glassy phenomena, eventually unlocking simulations beyond classical reach. Digital quantum simulations of glassy systems of realistic complexity, amenable through our proposed method, provide a framework for understanding the impact of disordered spin configurations on system dynamics and energy states. Extending on our methods could pave the way for simulating the glassy behavior of systems using quantum computers which are otherwise intractable. For future research, we propose exploring different phases in the Ising model across various geometries. Investigating how these competing fields interact in three-dimensional (3D) systems, lattice structures with higher coordination numbers, and more complex geometries could unlock new insights into the nature of disordered phases and the influence of field interactions. This could further elucidate the conditions under which glassy behavior emerges and how it can be controlled or manipulated. By exploring disorder effects in quantum systems, one can design catalysts with optimised distributions of active sites, enhancing both efficiency and selectivity. The ability to manipulate spin configurations and the resulting magnetic moment arrangements is pivotal in advancing spintronic devices and catalytic systems. The ability to simulate such complex phenomena on quantum hardware marks a significant step forward in leveraging quantum computational capabilities for condensed matter physics and material science. 

\section*{Acknowledgements}
We thank the anonymous referee for providing valuable suggestions which has helped us improve the presentation of our work. S.I.M. thanks Prof. Sabre Kais, Dr. Valentin Walther, Dr. Sumit Kale and Dr. Rishabh Gupta for their insightful discussions during the initial stages of this project. F.T.C. acknowledges support through the IBM Quantum Researchers Program for priority access to devices, and the Office of Naval Research for ﬁnancial support (ONR Award Number N62909-21-1-2018). The authors' opinions are their own, and they do not represent the official stance or policy of IBM. For the purpose of open access, the authors have applied a Creative Commons Attribution (CC BY) licence to any Author Accepted Manuscript version arising from this submission. 

\section*{Declarations}
\subsection*{Competing interests} 
The authors have no competing interests to report.
\subsection*{Data availability } 
The data that support the findings of this study are openly available in
\cite{das2024quglassyising}.

\subsection*{Code availability} 
The code that support the findings of this study are openly available in
\cite{das2024quglassyising}.

\appendix
\section{Appendix: Details of Real-Device Demonstration}

\subsection{IBM Q Oslo Device}
The real-device demonstration was conducted on the IBM Q Oslo quantum processor, a 7-qubit superconducting quantum device with a heavy-hexagonal qubit architecture. This topology limits each qubit's connectivity to two or three neighbors, reducing crosstalk errors and improving the reliability of computations. 

\subsection{IBM Q Oslo Qubit Layout and Quantum Circuit}
\begin{figure}[h]
  \centering
  \subfloat[IBM Q Oslo qubit connectivity layout.\label{fig:oslo_layout}]{%
    \includegraphics[width=0.45\textwidth]{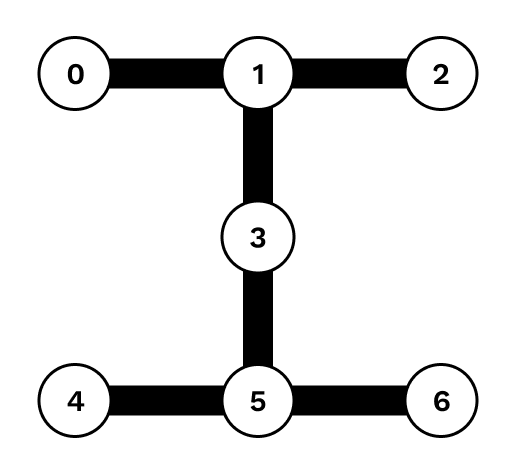}%
  }
  \hfill
  \subfloat[Quantum circuit executed on IBM Q Oslo.\label{fig:quantum_circuit}]{%
    \includegraphics[width=0.50\textwidth]{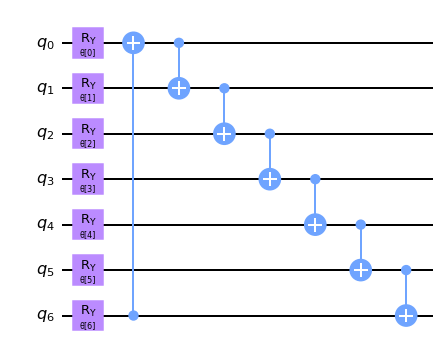}%
  }
  \caption{(a) Qubit layout of IBM Q Oslo \cite{Useche_2024}, illustrating the connectivity between qubits. (b) The quantum circuit implemented on the device, consisting of single-qubit $R_Y$ rotations and entangling gates.}
  \label{fig:overall}
\end{figure}

Fig.~\ref{fig:oslo_layout} shows the connectivity of the IBM Q Oslo qubits, where each node represents a qubit and the connections represent allowed two-qubit operations \cite{Useche_2024}. Fig.~\ref{fig:quantum_circuit} presents the quantum circuit used in real-device demonstration, consisting of single-qubit $R_Y$ rotations followed by a series of control operations implementing entanglement.

\subsection{Qiskit Runtime Environment}
The experiment was performed using the Qiskit Runtime environment, which allowed us to efficiently execute the quantum algorithms on the IBM Q Oslo device. The Variational Quantum Eigensolver (VQE) was used to compute the ground state of the 1D Ising Hamiltonian. The optimiser performed 46 cost function evaluations, leading to an eigenvalue of $-4.0139$ and an eigenstate distribution as detailed below.

\subsection{Results from the IBM Q Oslo Run}
The results from the quantum processor are as follows:
\begin{itemize}
    \item \textbf{Eigenvalue}: $-4.0139$
    \item \textbf{Optimal parameters}:
    \[
    \theta = \{4.3556, 3.2937, 2.8003, -3.1863, 4.0444, -2.9020, 3.9017\}
    \]
    \item \textbf{Optimal value}: $-4.0139$
    \item \textbf{optimiser evaluations}: 46
    \item \textbf{optimiser runtime}: $11743.05$ seconds
\end{itemize}

\noindent The eigenstate probabilities obtained from the run are shown in Table 1.

\begin{table}[h]
    \centering
    \renewcommand{\arraystretch}{1.2}
    \setlength{\tabcolsep}{6pt}
    \begin{tabular}{|c c||c c||c c|}
        \hline
        \textbf{State} & \textbf{Probability} & \textbf{State} & \textbf{Probability} & \textbf{State} & \textbf{Probability} \\
        \hline
        0000001 & 0.0316 & 0000100 & 0.0548 & 0000101 & 0.0632 \\
        0001001 & 0.0316 & 0001010 & 0.0775 & 0001011 & 0.0447 \\
        0001100 & 0.0316 & 0001101 & 0.0447 & 0001110 & 0.0316 \\
        0010000 & 0.0548 & 0010001 & 0.0548 & 0010010 & 0.0775 \\
        0010011 & 0.0548 & 0010100 & 0.1095 & 0010101 & 0.2191 \\
        0010110 & 0.0707 & 0011000 & 0.0316 & 0011001 & 0.0548 \\
        0011010 & 0.1183 & 0011011 & 0.0447 & 0011100 & 0.0632 \\
        0011101 & 0.0632 & 0011110 & 0.0316 & 0100001 & 0.0837 \\
        0100010 & 0.1549 & 0100100 & 0.1000 & 0100101 & 0.1949 \\
        0100110 & 0.1049 & 0101000 & 0.0894 & 0101001 & 0.1949 \\
        0101010 & 0.4506 & 0101011 & 0.1924 & 0101100 & 0.0707 \\
        0101101 & 0.1673 & 0101110 & 0.1140 & 0101111 & 0.0707 \\
        0110001 & 0.0316 & 0110010 & 0.0894 & 0110011 & 0.0316 \\
        0110100 & 0.0548 & 0110101 & 0.1095 & 0110110 & 0.0316 \\
        0110111 & 0.0316 & 0111000 & 0.0316 & 0111001 & 0.0316 \\
        0111010 & 0.1000 & 0111011 & 0.0707 & 0111100 & 0.0632 \\
        0111101 & 0.0632 & 0111110 & 0.0447 & 1000010 & 0.0707 \\
        1000100 & 0.0316 & 1000101 & 0.0894 & 1000110 & 0.0316 \\
        1000111 & 0.0316 & 1001000 & 0.0447 & 1001001 & 0.0316 \\
        1001010 & 0.1549 & 1001011 & 0.0837 & 1001100 & 0.0316 \\
        1001101 & 0.0548 & 1001110 & 0.0316 & 1010000 & 0.0548 \\
        1010001 & 0.0837 & 1010010 & 0.1673 & 1010011 & 0.0775 \\
        1010100 & 0.1673 & 1010101 & 0.2950 & 1010110 & 0.1549 \\
        1010111 & 0.0447 & 1011000 & 0.0548 & 1011001 & 0.0447 \\
        1011010 & 0.1643 & 1011100 & 0.0548 & 1011101 & 0.0548 \\
        1011110 & 0.0548 & 1100001 & 0.0447 & 1100010 & 0.0316 \\
        1100100 & 0.0632 & 1100101 & 0.1049 & 1100110 & 0.0316 \\
        1101000 & 0.0447 & 1101001 & 0.0775 & 1101010 & 0.1924 \\
        1101011 & 0.0837 & 1101100 & 0.0447 & 1101101 & 0.0894 \\
        1101110 & 0.0707 & 1101111 & 0.0316 & 1110010 & 0.0548 \\
        1110100 & 0.0316 & 1110101 & 0.0894 & 1111001 & 0.0316 \\
        1111010 & 0.0775 & 1111011 & 0.0316 & 1111101 & 0.0316 \\
        \hline
    \end{tabular}
    \caption{Eigenstate probabilities obtained from the IBM Q Oslo device.}
    \label{tab:eigenstates}
\end{table}





\bigskip
\bibliographystyle{sn-aps}
\bibliography{sn-bibliography}

\end{document}